\documentclass[11pt]{article}
\usepackage{amssymb,amsmath,amsfonts}
\usepackage{graphicx}
\usepackage{graphics}
\usepackage{eepic,epsfig}

\begin{document}

\title{Repulsive Casimir effect from extra dimensions and \\
Robin boundary conditions: from branes to pistons}
\author{E. Elizalde$^{1}$,\, S.~D. Odintsov$^{1,2}$\footnote{Also at Center of Theor. Phys., TSPU, Tomsk}, \, A.~A. Saharian$^{3,4}$
\\
%EndAName
\\
\textit{$^{1}$Instituto de Ciencias del Espacio (CSIC) }\\
\textit{and Institut d'Estudis Espacials de Catalunya (IEEC/CSIC) }\\
\textit{Campus UAB, Facultat de Ci\`{e}ncies, Torre C5-Parell-2a planta,}\\
\textit{08193 Bellaterra (Barcelona) Spain}\vspace{0.3cm}\\
\textit{$^2$Instituci\'{o} Catalana de Recerca i Estudis Avan\c{c}ats (ICREA)}\vspace{0.3cm}\\
\textit{$^3$Department of Physics, Yerevan State University,}\\
\textit{1 Alex Manoogian Street, 0025 Yerevan, Armenia}\vspace{0.3cm}\\
\textit{$^4$The Abdus Salam International Centre for Theoretical Physics,} \\
\textit{11 Strada Costiera, 34014 Trieste, Italy}} \maketitle

\begin{abstract}
We evaluate the Casimir energy and force for a massive scalar
field with general curvature coupling parameter, subject to Robin
boundary conditions on two codimension-one parallel plates,
located on a $(D+1)$-dimensional background spacetime with an
arbitrary internal space. The most general case of different Robin
coefficients on the two separate plates is considered. With
independence of the geometry of the internal space, the Casimir
forces are seen to be attractive for special cases of Dirichlet or
Neumann boundary conditions on both plates and repulsive for
Dirichlet boundary conditions on one plate and Neumann boundary
conditions on the other. For Robin boundary conditions, the
Casimir forces can be either attractive or repulsive, depending on
the Robin coefficients and the separation between the plates, what
is actually remarkable and useful. Indeed, we demonstrate the
existence of an equilibrium point for the interplate distance,
which is stabilized due to the Casimir force, and show that
stability is enhanced by the presence of the extra dimensions.
Applications of these properties in braneworld models are
discussed. Finally, the corresponding results are generalized to
the geometry of a piston of arbitrary cross section.
\end{abstract}

\bigskip


\bigskip

\section{Introduction}

\label{sec:introd}

Many of the high-energy theories of fundamental physics are
formulated in higher-dimensional spacetimes. In particular, the
idea of extra dimensions has been extensively used in supergravity
and superstring theories. It is commonly assumed that the extra
dimensions are compactified. From the inflationary point of view,
universes with compact spatial dimensions, under certain
conditions, should be considered a rule rather than an exception
\cite{Lind04}. Models involving a compact universe with
non-trivial topology play a very important role by providing
proper initial conditions for inflation. And compactification of
spatial dimensions leads to a number of interesting quantum field
theoretical effects, which include instabilities in interacting
field theories, topological mass generation, and symmetry
breaking.

In the case of non-trivial topology, the boundary conditions
imposed on fields give rise to a modification of the spectrum for
vacuum fluctuations and, as a result, to Casimir-type
contributions in the vacuum expectation values of physical
observables (for the topological Casimir effect and its role in
cosmology see \cite{Most97} and references therein). In models of
the Kaluza-Klein type, the Casimir effect has been used as a
stabilization mechanism for moduli fields and as a source for
dynamical compactification of the extra dimensions, in particular,
for quantum Kaluza-Klein gravity (see Ref. \cite{Buch89}). The
Casimir energy can also serve as a model for dark energy needed
for the explanation of the present accelerated expansion of the
universe (see \cite{Milt03} and references therein). In addition,
recent measurements of the Casimir forces between macroscopic
bodies provide a sensitive test for constraining the parameters of
long-range interactions, as predicted by modern unification
theories of fundamental interactions \cite{Most87}. The influence
of extra compactified dimensions on the Casimir effect in the
classical configuration of two parallel plates has been recently
discussed in \cite{Chen06}-\cite{Teo08}, for the case of a
massless scalar field with Dirichlet boundary conditions, and in
\cite{Popp04}-\cite{Peri08}, for the electromagnetic field for
perfectly conducting boundary conditions.

More recently, interest has concentrated on the topic of the
Casimir effect in braneworld models with large extra dimensions.
In this type of models (for a review see \cite{Brane}) the concept
of brane is used as a submanifold embedded in a higher dimensional
spacetime, on which the standard-model particles are confined.
Braneworlds naturally appear in the string/M theory context and
provide a novel set up for discussing phenomenological and
cosmological issues related with extra dimensions. In braneworld
models the investigation of quantum effects is of considerable
phenomenological interest, both in particle physics and in
cosmology. The braneworld corresponds to a manifold with
boundaries. All fields which propagate in the bulk will give
Casimir-type contributions to the vacuum energy and, as a result,
to the vacuum forces acting on the branes. Casimir forces provide
a natural mechanism for stabilizing the radion field in the
Randall-Sundrum model, as required for a complete solution of the
hierarchy problem. In addition, the Casimir energy gives a
contribution to both the brane and the bulk cosmological
constants. Hence, it has to be taken into account in any
self-consistent formulation of the braneworld dynamics. The
Casimir energy and corresponding Casimir forces within the
framework of the Randall-Sundrum braneworld \cite{Rand99} have
been evaluated in Refs.~\cite{Gold00}-\cite{Fran07} by using both
dimensional and zeta function regularization methods. Local
Casimir densities were considered in Refs.~\cite{Knap04,Saha05}.
The Casimir effect in higher dimensional generalizations of the
Randall-Sundrum model with compact internal spaces has been
investigated in \cite{Flac03}-\cite{Fran08}.

The purpose of the present paper is to study the Casimir energy
and force for a massive scalar field with an arbitrary curvature
coupling parameter, obeying Robin boundary conditions on two
codimension one parallel plates which are embedded in the
background spacetime $R^{(D_{1}-1,1)}\times \Sigma $, being
$\Sigma $ an arbitrary compact internal space. The most general
case is considered, where the constants in the boundary conditions
are different for the two separate plates. It will be shown that
Robin boundary conditions with different coefficients are
necessary to obtain repulsive Casimir forces. Robin type
conditions are an extension of Dirichlet and Neumann boundary
conditions and genuinely appear in a variety of situations,
including vacuum effects for a confined charged scalar field in
external fields \cite{Ambj83}, spinor and gauge field theories,
quantum gravity and supergravity \cite{Luck91}. Robin conditions
can be made conformally invariant, while purely-Neumann conditions
cannot. Therefore, Robin type conditions are needed when one deals
with conformally invariant theories in the presence of boundaries
and wishes to preserve this invariance. It is interesting to note
that a quantum scalar field satisfying Robin conditions on the
boundary of a cavity violates the Bekenstein's entropy-to-energy
bound near certain points in the space of the parameter defining
the boundary conditions \cite{Solo01}. Robin boundary conditions
are an extension of those imposed on perfectly conducting
boundaries and may, in some geometries, be useful for modelling
the finite penetration of the field through the boundary, the
skin-depth parameter being related to the Robin coefficient
\cite{Most85,Lebe01}. In other words, those are the boundary
conditions which are more suitable to describe physically
realistic situations. This type of boundary conditions naturally
arise for scalar and fermion bulk fields in the Randall-Sundrum
model \cite{Flac01,Saha05,Gher00} and the corresponding Robin
coefficients are related to the curvature scale and to the
boundary mass terms of the field. Robin boundary conditions also
appear in the study of Casimir forces between the boundary planes
of films (for a recent discussion with references see, for
instance, \cite{Schm08}). The Casimir effect in the geometry of
two parallel plates with Robin boundary condition was investigated
in Refs.~\cite{Lebe01,Rome02,Full03,Albu04,Bajn06}. Note moreover
that boundary problems with non-local boundary conditions can also
be reduced to corresponding ones with Robin conditions, with the
coefficients depending on the wave vector components along the
plates~\cite{Saha05b}.

The outline of the paper is as follows. In the next section we
will consider the geometry of the problem and the corresponding
eigenfunctions. The Casimir energy for two parallel plates in the
general case for the internal subspace is evaluated in
Sect.~\ref{sec:CasEn}. The boundary-free and single plate parts
will be extracted in a cutoff independent way. Applications to
braneworlds are then discussed. In Sect.~\ref{sec:CasForce} we
consider the Casimir forces and show that depending on the
coefficients in the boundary conditions these forces can be either
attractive or repulsive. The asymptotic behavior of the forces,
for small and large interplate distances, is given. As an
application of the general results, in Sect.~\ref{sec:example} a
simple example with an internal space $S^{1}$ is discussed, for
general periodicity condition along the compactified dimension.
For this special example, we also present the boundary-free part
and extract from the single plate parts the topological
contributions. The corresponding generalizations for the internal
spaces $(S^1)^N$ and $S^{N}$ are also given in detail. In
Sect.~\ref{sec:Piston} we extend the results for the Casimir
energy and force to the case of the geometry of a piston with
arbitrary cross-section. Section \ref{sec:Conc} contains a summary
of the work.

\section{Geometry of the problem and eigenfunctions}

\label{sec:Geometry}

We consider a scalar field $\varphi (x)$, with arbitrary curvature coupling parameter
$\zeta $, satisfying the equation of motion
\begin{equation}
\left( g^{MN}\nabla _{M}\nabla _{N}+m^{2}+\zeta R\right) \varphi (x)=0,
\label{fieldeq}
\end{equation}%
$M,N=0,1,\ldots ,D$, with $R$ being the scalar curvature for a $(D+1)$%
-dimensional background spacetime (for the metric signature and the
curvature tensor we adopt the conventions of Ref. \cite{Birr82}). For the special
cases of minimally and of conformally coupled scalars one has, respectively,
$\zeta =0$ and $\zeta =\zeta _{D}\equiv (D-1)/4D$. We will assume that the
background spacetime has a topology $R^{(D_{1},1)}\times \Sigma $, where $%
R^{(D_{1},1)}$ is $(D_{1}+1)$-dimensional Minkowski spacetime and
$\Sigma $ a $D_{2}$-dimensional internal manifold,
$D=D_{1}+D_{2}$. The corresponding line element has the form
\begin{equation}
ds^{2}=g_{MN}dx^{M}dx^{N}=\eta _{\mu \nu }dx^{\mu }dx^{\nu }-\gamma
_{il}dX^{i}dX^{l},  \label{metric}
\end{equation}%
with $\eta _{\mu \nu }=\mathrm{diag}(1,-1,\ldots ,-1)$ being the metric for
the $(D_{1}+1)$-dimensional Minkowski spacetime and the coordinates $X^{i}$
cover the manifold $\Sigma $. Here and below $\mu ,\nu
=0,1,\ldots ,D_{1}$ and $i,l=1,\ldots ,D_{2}$. For the scalar curvature of
the metric tensor, from (\ref{metric}) one has $R=-R_{(\gamma )}$, where $%
R_{(\gamma )}$ is the scalar curvature for the metric tensor $\gamma _{il}$.

Our main interests in this paper will be to study the Casimir energy
density and the mutual forces  occurring for
the geometry of two parallel infinite plates of codimension one, located at $%
x^{D_{1}}=a_{1}$ and $x^{D_{1}}=a_{2}$, $a_{1}<a_{2}$. As most
general set up, we assume that on these boundaries the scalar
field obeys Robin boundary conditions
\begin{equation}
\left( 1+\beta _{j}n^{M}\nabla _{M}\right) \varphi (x)=[1+\beta
_{j}(-1)^{j-1}\partial _{D_{1}}]\varphi (x)=0,\quad x^{D_{1}}=a_{j},\;j=1,2,
\label{boundcond}
\end{equation}%
with constant coefficients $\beta _{j}$. For $\beta _{j}=0$ these
boundary conditions are reduced to Dirichlet one and for $\beta
_{j}=\infty $ to Neumann boundary conditions. The choice of
different boundary conditions on the plates may correspond
physically to use of different materials for plates. The
imposition of boundary conditions on the quantum field changes the
spectrum for the zero--point fluctuations and leads to the
modification of the vacuum expectation values for physical
quantities, as compared with the same situation without
boundaries.

In the region between the plates, $a_{1}<x^{D_{1}}<a_{2}$, the corresponding
eigenfunctions, satisfying the boundary condition on the plate at $%
x^{D_{1}}=a_{j}$, can be expressed in the decomposed form:
\begin{equation}
\varphi _{\alpha }(x^{M})=C_{\alpha }\exp \left(-i\sum_{\mu ,\nu
=0}^{D_{1}-1}\eta _{\mu \nu }k^{\mu }x^{\nu }\right)\cos
\left[k^{D_{1}}|x^{D_{1}}-a_{j}|+\alpha _{j}\right]\, \psi _{\beta }(X),  \label{eigfunc1}
\end{equation}%
where $\alpha $ denotes a set of quantum numbers specifying the solution and
\begin{eqnarray}
&& k^{0}=\omega =\sqrt{k^{2}+\left( k^{D_{1}}\right) ^{2}+m_{\beta }^{2}}%
,\quad m_{\beta }^{2}=\lambda _{\beta }^{2}+m^{2},  \notag \\
&& k =|\mathbf{k}|,\;\mathbf{k}=(k^{1},\ldots ,k^{D_{1}-1}).  \label{mbet}
\end{eqnarray}%
In Eq. (\ref{eigfunc1}), the $\alpha _{j}$, $j=1,2$, are defined by the
relations
\begin{equation}
\sin \alpha _{j}=\frac{1}{\sqrt{\left( k^{D_{1}}\right) ^{2}\beta _{j}^{2}+1}%
},\;\cos \alpha _{j}=\frac{k^{D_{1}}\beta _{j}}{\sqrt{\left(
k^{D_{1}}\right) ^{2}\beta _{j}^{2}+1}}.  \label{sinalfaj}
\end{equation}
The modes $\psi _{\beta }(X)$ are the eigenfunctions of the operator $%
\Delta _{(\gamma )}+\zeta R_{(\gamma )}$:
\begin{equation}
\left[ \Delta _{(\gamma )}+\zeta R_{(\gamma )}\right] \psi _{\beta
}(X)=-\lambda _{\beta }^{2}\psi _{\beta }(X),  \label{eqint1}
\end{equation}%
with eigenvalues $\lambda _{\beta }^{2}$, and fulfill the
normalization condition
\begin{equation}
\int d^{D_{2}}X\,\sqrt{\gamma }\psi _{\beta }(X)\psi _{\beta ^{\prime
}}^{\ast }(X)=\delta _{\beta \beta ^{\prime }}.  \label{normpsibet}
\end{equation}%
In Eq. (\ref{eqint1}), $\Delta _{(\gamma )}$ is the Laplace-Beltrami
operator for the metric $\gamma _{il}$. In the consideration below we will
assume that $\lambda _{\beta }\geqslant 0$.

From the boundary condition on the second plate one obtains that the
eigenvalues for $k^{D_{1}}$ are solutions of the equation%
\begin{eqnarray}
F(z) &=&(1-b_{1}b_{2}z^{2})\sin z-(b_{1}+b_{2})z\cos z=0,\;  \notag \\
z &=&ak^{D_{1}},\;a=a_{2}-a_{1},\;b_{j}=\beta _{j}/a.  \label{eigeq}
\end{eqnarray}%
We denote by $z=z_{n}$, $n=1,2,\ldots $, the zeros of the function $F(z)$ in
the right half-plane of the complex variable $z$, arranged in  ascending
order, $z_{n}<z_{n+1}$. In the discussion below we will assume that all
these zeros are real. This is the case for the conditions (see \cite%
{Rome02}) $\{b_{1}+b_{2}\geqslant 1,b_{1}b_{2}\leqslant 0\}\cup
\{b_{1,2}\leqslant 0\}$. The coefficient $C_{\alpha }$ in (\ref{eigfunc1})
is determined from the orthonormality condition for the eigenfunctions, and
is equal to%
\begin{equation}
C_{\alpha }^{2}=\frac{(2\pi )^{1-D_{1}}}{\omega (z_{n})a}\left[ 1+\frac{1}{%
z_{n}}\sin (z_{n})\cos (z_{n}+2\alpha _{j})\right] ^{-1},  \label{C2}
\end{equation}%
being $\omega (z_{n})=\sqrt{k^{2}+z_{n}^{2}/a^{2}+m_{%
\beta }^{2}}$ the eigenfrequencies.

\section{The Casimir energy}

\label{sec:CasEn}

The vacuum energy in the region between the plates (per unit volume along
the directions $x^{1},\ldots ,x^{D_{1}-1}$) is given by the formal expression%
\begin{equation}
E_{[a_{1},a_{2}]}=\frac{1}{2}\int \frac{d\mathbf{k}}{(2\pi )^{D_{1}-1}}%
\sum_{\beta }\sum_{n=1}^{\infty }\sqrt{k^{2}+z_{n}^{2}/a^{2}+m_{\beta }^{2}}.
\label{ELR}
\end{equation}%
In the discussion below we will assume that some cutoff function is present,
without writing it explicitly. Alternatively, one can use zeta function
regularization, that yields the same result. For the sum over $n$ we use
the summation formula
\cite{Rome02,SahRev07}%
\begin{eqnarray}
\sum_{n=1}^{\infty }\frac{\pi f(z_{n})}{1+\sin (z_{n})\cos (z_{n}+2\alpha
_{j})/z_{n}} &=&-\frac{\pi }{2}\frac{f(0)}{1-b_{2}-b_{1}}+\int_{0}^{\infty
}dzf(z)  \notag \\
&&+i\int_{0}^{\infty }dz\frac{f(iz)-f(-iz)}{\frac{(b_{1}z-1)(b_{2}z-1)}{%
(b_{1}z+1)(b_{2}z+1)}e^{2z}-1}.  \label{sumfor}
\end{eqnarray}%
By taking into account the relation%
\begin{equation}
1+\frac{\sin (z_{n})}{z_{n}}\cos (z_{n}+2\alpha _{j})=1-\sum_{j=1}^{2}\frac{%
b_{j}}{1+b_{j}^{2}z_{n}^{2}},  \label{rel1}
\end{equation}%
we see that the sum on the left-hand side of (\ref{sumfor}) coincides with
the corresponding sum in the vacuum energy, if we take%
\begin{equation}
f(z)=\sqrt{k^{2}+z^{2}/a^{2}+m _{\beta }^{2}}\left( 1-\sum_{j=1}^{2}\frac{%
b_{j}}{1+b_{j}^{2}z^{2}}\right) .  \label{finAPF}
\end{equation}

The application of the summation formula (\ref{sumfor}) with (\ref{finAPF})
allows us to write the vacuum energy from (\ref{ELR}) in the decomposed
form%
\begin{equation}
E_{[a_{1},a_{2}]}=aE_{R^{(D_{1},1)}\times \Sigma }+\sum_{j=1,2}E_{j}+\Delta
E_{[a_{1},a_{2}]},  \label{Edec}
\end{equation}%
where we have introduced the notations%
\begin{equation}
E_{j}=-\frac{1}{8}\int \frac{d\mathbf{k}}{(2\pi )^{D_{1}-1}}\sum_{\beta }%
\sqrt{k^{2}+m_{\beta }^{2}}-\frac{\beta _{j}}{2\pi }\int \frac{d\mathbf{k}}{%
(2\pi )^{D_{1}-1}}\sum_{\beta }\int_{0}^{\infty }dx\,\frac{\sqrt{%
k^{2}+x^{2}+m_{\beta }^{2}}}{1+\beta _{j}^{2}x^{2}},  \label{Ej}
\end{equation}%
and
\begin{equation}
\Delta E_{[a_{1},a_{2}]}=-\frac{1}{\pi }\int \frac{d\mathbf{k}}{(2\pi
)^{D_{1}-1}}\sum_{\beta }\int_{\sqrt{k^{2}+m_{\beta }^{2}}}^{\infty }dz\frac{%
\sqrt{z^{2}-k^{2}-m_{\beta }^{2}}}{\frac{(\beta _{1}z-1)(\beta _{2}z-1)}{%
(\beta _{1}z+1)(\beta _{2}z+1)}e^{2az}-1}\left( a+\sum_{j=1}^{2}\frac{\beta
_{j}}{\beta _{j}^{2}z^{2}-1}\right) .  \label{DelE0}
\end{equation}%
In Eq.~(\ref{Edec}),
\begin{equation}
E_{R^{(D_{1},1)}\times \Sigma }=\frac{1}{2}\int \frac{d\mathbf{k}_{D_{1}}}{%
(2\pi )^{D_{1}}}\sum_{\beta }\sqrt{k_{D_{1}}^{2}+m_{\beta }^{2}}
\label{Etop}
\end{equation}%
is the vacuum energy (per unit volume along the directions $x^{1},\ldots
,x^{D_{1}}$) in the spacetime of topology $R^{(D_{1},1)}\times \Sigma $
for the case
when the plates are absent. In the limit $a\rightarrow \infty $ the term $%
\Delta E_{[a_{1},a_{2}]}$ vanishes and the contribution $E_{j}$\ can be interpreted
as the vacuum energy (per unit volume along the directions $x^{1},\ldots
,x^{D_{1}-1}$) induced by the presence of the plate located at $%
x^{D_{1}}=a_{j}$ in the half-space $x^{D_{1}}\geqslant a_{j}$. These single
plate components do not depend on the location of the plate and do not contribute
to the vacuum force acting on the plates. As it will be shown below, the
latter is determined by the term $\Delta E_{[a_{1},a_{2}]}$. Note that this
contribution is finite and that the cutoff function is strictly necessary for the terms $%
E_{R^{(D_{1},1)}\times \Sigma }$ and $E_{j}$ only. In the discussion below
we will refer to $\Delta E_{[a_{1},a_{2}]}$ as the interaction term.

For further simplification of the corresponding expression, we use the
relation%
\begin{equation}
\int \frac{d\mathbf{k}}{(2\pi )^{D_{1}-1}}\int_{\sqrt{k^{2}+m_{\beta }^{2}}%
}^{\infty }dz\sqrt{z^{2}-k^{2}-m_{\beta }^{2}}g(z)=\frac{(4\pi )^{1-D_{1}/2}%
}{2D_{1}\Gamma (D_{1}/2)}\int_{m_{\beta }}^{\infty }dx\,(x^{2}-m_{\beta
}^{2})\,^{D_{1}/2}g(x).  \label{rel3}
\end{equation}%
In order to derive this formula, we must first integrate the left-hand side
over the angular part of the vector $\mathbf{k}$ and then change to
a new integration variable, $y=\sqrt{z^{2}-k^{2}-m_{\beta }^{2}}$. After
introducing polar coordinates in the $(k,y)$-plane and integrating
over the polar angle, we get Eq. (\ref{rel3}). By using this relation,
for the interaction part of the vacuum energy we find
\begin{equation}
\Delta E_{[a_{1},a_{2}]}=-\frac{(4\pi )^{-D_{1}/2}}{\Gamma (D_{1}/2+1)}%
\sum_{\beta }\int_{m_{\beta }}^{\infty }dx\,\frac{(x^{2}-m_{\beta
}^{2})\,^{D_{1}/2}}{\frac{(\beta _{1}x-1)(\beta _{2}x-1)}{(\beta
_{1}x+1)(\beta _{2}x+1)}e^{2ax}-1}\left( a+\sum_{j=1}^{2}\frac{\beta _{j}}{%
\beta _{j}^{2}x^{2}-1}\right) .  \label{DelE2}
\end{equation}%
Using
\begin{equation}
\left( a+\sum_{j=1}^{2}\frac{\beta _{j}}{\beta _{j}^{2}x^{2}-1}\right) \frac{%
2}{\frac{(\beta _{1}x-1)(\beta _{2}x-1)}{(\beta _{1}x+1)(\beta _{2}x+1)}%
e^{2ax}-1}=\frac{d}{dx}\ln \left[ 1-\frac{(\beta _{1}x-1)(\beta _{2}x-1)}{%
(\beta _{1}x+1)(\beta _{2}x+1)}e^{-2ax}\right] ,  \label{rel2}
\end{equation}%
and integrating by parts, the interaction term in the vacuum
energy can be written as
\begin{eqnarray}
\Delta E_{[a_{1},a_{2}]} &=&\frac{(4\pi )^{-D_{1}/2}}{\Gamma (D_{1}/2)}%
\sum_{\beta }\int_{m_{\beta }}^{\infty }dx\,x(x^{2}-m_{\beta
}^{2})\,^{D_{1}/2-1}  \notag \\
&&\times \ln \left[ 1-\frac{(\beta _{1}x+1)(\beta _{2}x+1)}{(\beta
_{1}x-1)(\beta _{2}x-1)}e^{-2ax}\right] \,.  \label{DelE}
\end{eqnarray}%
In the case when the internal space is absent and for a massless scalar
field this result reduces to the one derived in \cite{Rome02}. Note that the
bulk divergences in the vacuum energy between the plates are contained in
the first term on the right-hand side of (\ref{Edec}) and the boundary
divergences are contained in the single plate contributions $E_{j}$. The interaction
part is unambiguously defined. In particular, it does not depend on the
regularization scheme used (see, for example, Ref.~\cite{Rome02} for the
case without the internal space, where exactly the same result is obtained with
zeta function techniques).

For the special cases of Dirichlet and Neumann boundary conditions on both
plates, from (\ref{DelE}) one finds%
\begin{equation}
\Delta E_{[a_{1},a_{2}]}^{\mathrm{(J,J)}}=-\frac{(4\pi )^{-D_{1}/2}a}{\Gamma
(D_{1}/2+1)}\sum_{\beta }\int_{m_{\beta }}^{\infty }dx\,\frac{%
(x^{2}-m_{\beta }^{2})\,^{D_{1}/2}}{e^{2ax}-1},  \label{DelEDN}
\end{equation}%
with $\mathrm{J=D}$ and $\mathrm{J=N}$ for Dirichlet and Neumann boundary
conditions, respectively. By expanding the factor $1/(e^{2ax}-1)$ in the
integrand one gets
\begin{equation}
\int_{m_{\beta }}^{\infty }dx\,\frac{(x^{2}-m_{\beta }^{2})\,^{D_{1}/2}}{%
e^{2ax}-1}=\frac{\Gamma (D_{1}/2+1)}{\sqrt{\pi }a^{D_{1}+1}}%
\sum_{n=1}^{\infty }\left( \frac{am_{\beta }}{n}\right)
^{(D_{1}+1)/2}K_{(D_{1}+1)/2}(2nam_{\beta }),  \label{rel4}
\end{equation}%
being $K_{\nu }(z)$ the Mac-Donald (or modified Bessel) function. This allows us to write
the corresponding vacuum energy for Dirichlet and Neumann scalars as
\begin{equation}
\Delta E_{[a_{1},a_{2}]}^{\mathrm{(J,J)}}=-\frac{2a^{-D_{1}}}{(8\pi
)^{(D_{1}+1)/2}}\sum_{\beta }\sum_{n=1}^{\infty }\frac{%
f_{(D_{1}+1)/2}(2nam_{\beta })}{n^{D_{1}+1}},  \label{DELEDN1}
\end{equation}%
with the notation%
\begin{equation}
f_{\nu }(z)=z^{\nu }K_{\nu }(z).  \label{fnu}
\end{equation}%
The energy given by Eq.~(\ref{DELEDN1}) is always negative and the
corresponding Casimir forces are attractive for
all interplate distances, as will be shown below.
In the case $D_{1}=3$ and for a massless scalar
field, Eq.~(\ref{DELEDN1}) reduces to the expression given in Ref.~\cite%
{Full09b}, where the zeta function method was used.

For Dirichlet boundary conditions on one plate and Neumann boundary conditions
on the other, similarly to (\ref{DelE2}) we get%
\begin{eqnarray}
\Delta E_{[a_{1},a_{2}]}^{\mathrm{(D,N)}} &=&\frac{(4\pi )^{-D_{1}/2}a}{%
\Gamma (D_{1}/2+1)}\sum_{\beta }\int_{m_{\beta }}^{\infty }dx\,\frac{%
(x^{2}-m_{\beta }^{2})\,^{D_{1}/2}}{e^{2ax}+1}  \notag \\
&=&-\frac{2a^{-D_{1}}}{(8\pi )^{(D_{1}+1)/2}}\sum_{\beta }\sum_{n=1}^{\infty
}\frac{f_{(D_{1}+1)/2}(2nam_{\beta })}{(-1)^{n}n^{D_{1}+1}}.  \label{DelED+N}
\end{eqnarray}%
In this case the energy $\Delta E_{[a_{1},a_{2}]}$ is always positive and
the corresponding vacuum forces are repulsive for all distances between the
plates.

By using the result (\ref{DelE}), in a similar way as in
\cite{Saha04}, we obtain the corresponding Casimir energy for a
conformally coupled massless scalar field $\overline{\varphi }(x)$ on the
background of a spacetime with metric tensor $\overline{g}_{MN}=\Omega
^{2}(x^{D_{1}})g_{MN}$, where the metric $g_{MN}$ is defined by the line element
(\ref{metric}). We assume that the field obeys the boundary conditions:
\begin{equation}
(1+\overline{\beta }_{j}n^{M}\nabla _{M})\overline{\varphi }%
(x)=[1+(-1)^{j-1}\Omega _{j}^{-1}\overline{\beta }_{j}\partial _{D_{1}}]%
\overline{\varphi }(x)=0,\;\Omega _{j}=\Omega (x_{j}^{D_{1}}),
\label{BCbrane}
\end{equation}%
on two codimension-one branes with coordinates $x^{D_{1}}=a_{j}$,$\;j=1,2$.
The corresponding results for the interaction part of the Casimir energy can
be derived from those obtained before simply by using the conformal
relation that relates the two problems. The fields are connected by the formula $%
\overline{\varphi }(x)=\Omega ^{(1-D)/2}\varphi (x)$. Making use of this
relation, from Eqs.~(\ref{boundcond}) and (\ref{BCbrane}) we obtain the
following relations between the Robin coefficients:%
\begin{equation}
\beta _{j}=\left[ \Omega _{j}+(-1)^{j}\frac{D-1}{2\Omega _{j}}\overline{%
\beta }_{j}\Omega _{j}^{\prime }\right] ^{-1}\overline{\beta }_{j},
\label{relbeta}
\end{equation}%
where $\Omega _{j}^{\prime }=\Omega _{j}^{\prime }(x_{j}^{D_{1}})$. We
conclude that for a conformally coupled massless scalar field with boundary
conditions (\ref{BCbrane}), the interaction part of the vacuum energy in the
region between the branes is given by  (\ref{DelE}), where the
coefficients $\beta _{j}$ are defined by the relations (\ref{relbeta}). In
particular, for the case of Neumann boundary conditions ($1/\overline{\beta }%
_{j}=0$), one has $\beta _{j}=2\Omega _{j}(-1)^{j}/[(D-1)\Omega _{j}^{\prime
}]$. In the special case of the AdS bulk used in the Randall-Sundrum braneworld
model \cite{Rand99} (note that in this model only the inside region between the branes
is considered) we have $\Omega (x^{D_{1}})=r_{D}/x^{D_{1}}$, being $r_{D}$
the AdS radius. The corresponding Robin coefficients for an untwisted
scalar are given by the relations \cite{Flac01,Saha05,Saha06,Gher00}
\begin{equation}
\overline{\beta }_{j}^{-1}=(-1)^{j}c_{j}/2-2D\zeta /r_{D},  \label{betajRS}
\end{equation}%
where $c_{1}$ and $c_{2}$ are the mass parameters in the surface action of
the scalar field for the left and right branes, respectively. For a twisted
scalar field, Dirichlet boundary conditions are obtained on both branes.

To summarize, as we see, in the case of the warped geometry the corresponding vacuum
energy is not, in general, a monotonic function of the inter-brane distance
and can display a minimum, corresponding to the stable equilibrium point. This
property can be used in braneworld models for the stabilization of the radion field.
An important difference between the warped geometry and the one discussed before
is that now the single brane contributions to the vacuum energy depend on the
location of the brane and, hence, give additional contributions to the force
acting on the brane. The divergences in the single brane components are absorbed
by adding the respective counterterms to the brane action. The coefficients
of these counterterms are not computable within the framework of the
low-energy effective theory and should be considered as parameters which are
fixed by imposing renormalization conditions on the corresponding effective
potential (see also the discussions in Refs.~\cite%
{Gold00,Flac01,Garr01,Flac03,Flac03b,Saha06b,Pont01}).

\section{The Casimir force}

\label{sec:CasForce}

The vacuum energy corresponding to the region $0\leqslant x^{l}\leqslant c_{l}$, $%
l=1,\ldots ,D_{1}-1$, $a_{1}\leqslant x^{D_{1}}\leqslant a_{2}$ will be denoted $%
E_{[a_{1},a_{2}]}c_{1}\cdots c_{D_{1}-1}$. The volume of this region is $%
V=V_{\Sigma }c_{1}\cdots c_{D_{1}-1}a$, being $V_{\Sigma }$ the volume
of the internal space. The corresponding vacuum stress at $x^{D_{1}}=a_{1}+$
is given by%
\begin{equation}
P(a_{1}+)=-\frac{\partial }{\partial V}E_{[a_{1},a_{2}]}c_{1}\cdots
c_{D_{1}-1}=P_{0}+\Delta P(a_{1}+),  \label{P}
\end{equation}%
where%
\begin{equation}
P_{0}=-\frac{E_{R^{(D_{1},1)}\times \Sigma }}{V_{\Sigma }},\;\Delta
P(a_{1}+)=-\frac{1}{V_{\Sigma }}\frac{\partial }{\partial a}\Delta
E_{[a_{1},a_{2}]}.  \label{P0}
\end{equation}%
Using Eq.~(\ref{DelE}), we find
\begin{equation}
\Delta P(a_{1}+)=-\frac{2(4\pi )^{-D_{1}/2}}{V_{\Sigma }\Gamma (D_{1}/2)}%
\sum_{\beta }\int_{m_{\beta }}^{\infty }dx\,x^{2}\frac{(x^{2}-m_{\beta
}^{2})\,^{D_{1}/2-1}}{\frac{(\beta _{1}x-1)(\beta _{2}x-1)}{(\beta
_{1}x+1)(\beta _{2}x+1)}e^{2ax}-1}.  \label{P1}
\end{equation}%
As can be easily seen, the vacuum stress at $x^{D_{1}}=a_{2}-$ is given
by the same expression: $\Delta P(a_{2}-)=\Delta P(a_{1}+)$.

For the geometry of two parallel plates, the total vacuum energy is the sum
of the contributions from the regions $x^{D_{1}}\leqslant a_{1}$, $%
a_{1}\leqslant x^{D_{1}}\leqslant a_{2}$ and $a_{2}\leqslant x^{D_{1}}$. When
investigating the resulting force on the plate at $x^{D_{1}}=a_{1}$, in
order to deal with finite regions from both sides, we will consider a piston
like geometry (with large transverse dimensions, for a piston with finite
cross section see below), assuming the presence of an additional plate
located at $x^{D_{1}}=a_{0}<a_{1}$. For the corresponding vacuum stress at $%
x^{D_{1}}=a_{1}-$, one has%
\begin{equation}
P(a_{1}-)=P_{0}+\Delta P(a_{1}-),\;\Delta P(a_{1}-)=-\frac{1}{V_{\Sigma }}%
\frac{\partial }{\partial b}\Delta E_{[b_{1},a_{1}]},  \label{P1-}
\end{equation}%
with $b=a_{1}-a_{0}$. The resulting pressure on the plate at $x^{D_{1}}=a_{1}
$ is given by the difference%
\begin{equation}
P(a_{1})=\Delta P(a_{1}+)-\Delta P(a_{1}-).  \label{Pa1}
\end{equation}%
As we see, the contributions to the vacuum force coming from the term $P_{0}$ are the same from
the left and from the right sides of the plate, so that there is no netto contribution to the effective
force. In the limit $a_{0}\rightarrow -\infty $ one has $\Delta
P(a_{1}-)\rightarrow 0$ and the Casimir force acting on the plate at $%
x^{D_{1}}=a_{j}$, $j=1,2$, in the original two-plate geometry is given
by the expression
\begin{equation}
P=-\frac{2(4\pi )^{-D_{1}/2}}{V_{\Sigma }\Gamma (D_{1}/2)a^{D_{1}+1}}%
\sum_{\beta }\int_{am_{\beta }}^{\infty }dx\,\frac{x^{2}(x^{2}-a^{2}m_{\beta
}^{2})\,^{D_{1}/2-1}}{\frac{(b_{1}x-1)(b_{2}x-1)}{(b_{1}x+1)(b_{2}x+1)}%
e^{2x}-1}.  \label{PCas}
\end{equation}%
This force is attractive when $P<0$ and repulsive when $P>0$. If one does
not take into account the contributions from the exterior regions $x^{D_{1}}\leqslant
a_{1}$ and $x^{D_{1}}\geqslant a_{2}$, the effective pressure is given by (%
\ref{P}) where the renormalized value for $P_{0}$ does not depend on the
separation of the plates. In the case $\Sigma =S^{1}$ and for periodic boundary
conditions along the compactified dimension, the renormalized value $P_{0}$
is positive (see next section, Eq.~(\ref{EL2})) which would correspond to
the repulsive force between the plates observed in the first paper of \cite%
{Chen06} (see also the discussion in Ref.~\cite{Full09}).

\begin{figure}[tbph]
\begin{center}
\epsfig{figure=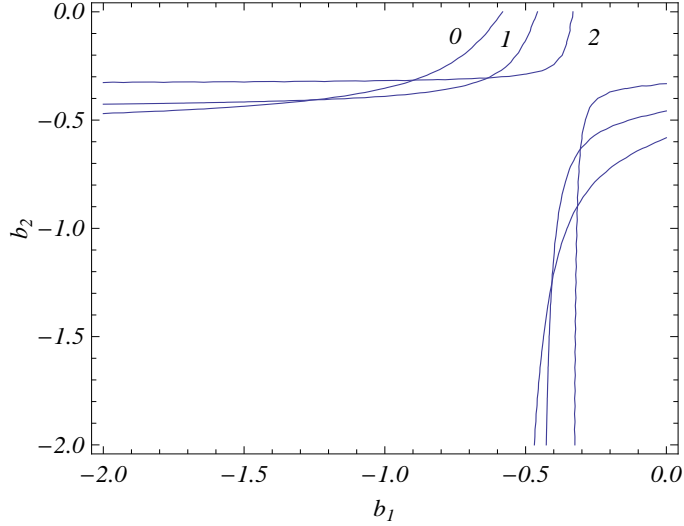,width=9.cm,height=7.cm}
\end{center}
\caption{The location of the zeros for the integral in (\protect\ref{PCas})
in the case $D_{1}=3$ and for different values of the parameter $am_{\protect%
\beta }$ (numbers near the curves).}
\label{fig1}
\end{figure}

As is clearly seen from Eq.~(\ref{PCas}), the sign of the vacuum stress on
the plate is determined by the sign of the integral in this formula. In
Fig.~\ref{fig1} we have plotted the location of the zeros for this
integral on the $(b_{1},b_{2})$-plane in the case $D_{1}=3$ and for
different values of $am_{\beta }$ (figures on the curves). For a given $%
am_{\beta }$, these zeros are located on two curves symmetric with respect
to the line $b_{1}=b_{2}$. The integral is positive in the region containing
this line and it is negative outside. In particular, the Casimir force between
the plates is always attractive for symmetric boundary conditions with $%
\beta _{1}=\beta _{2}<0$. This result is a special case of the general
theorem \cite{Kenn06}, which dictates an attraction between bodies with the {\it same}
properties. Note that the curves in Fig.~\ref{fig1} display the locations
of the zeros for the Casimir force in the geometry of two parallel plates on the
background of a 4-dimensional Minkowski spacetime.

For the special cases of Dirichlet and Neumann boundary conditions,
making use of the recurrence relations for the function $K_{\nu }(z)$, the
Casimir forces can be written as
\begin{eqnarray}
P^{\mathrm{(J,J)}} &=&-\frac{2(4\pi )^{-D_{1}/2}}{V_{\Sigma }\Gamma (D_{1}/2)%
}\sum_{\beta }\int_{m_{\beta }}^{\infty }dx\,x^{2}\frac{(x^{2}-m_{\beta
}^{2})\,^{D_{1}/2-1}}{e^{2ax}-1}  \notag \\
&=&\frac{2a^{-D_{1}-1}}{(8\pi )^{(D_{1}+1)/2}V_{\Sigma }}\sum_{\beta
}\sum_{n=1}^{\infty }\frac{1}{n^{D_{1}+1}}\left[ f_{(D_{1}+1)/2}(2nam_{\beta
})-f_{(D_{1}+3)/2}(2nam_{\beta })\right] ,  \label{FDN}
\end{eqnarray}%
with $\mathrm{J=D,N}$. Again, in the case $D_{1}=3$ and for a massless
scalar field this result reduces to the one obtained in \cite%
{Full09b}. The forces described by Eq.~(\ref{FDN}) are attractive for all
distances between the plates, irrespective of the geometry of the internal
subspace. For Dirichlet boundary conditions on one plate and Neumann boundary
conditions on the other, the expression for the Casimir force is obtained from
Eq.~(\ref{FDN}) after introducing an additional factor $(-1)^{n+1}$ in the
summation over $n$. In this case,  Casimir forces are repulsive. In the
general case of Robin boundary conditions the Casimir forces can be either
attractive or repulsive, depending on the coefficients present
in the definition of the boundary
conditions, and on the distance between the plates. For the special case of the
topology $R^{(D-1,1)}\times S^{1}$ this issue will be illustrated in the next section.

Let us now consider the asymptotic behavior of the Casimir force as a
function of the size of the internal space. Note that if the size of the
internal space is of the order $L$, then for nonzero modes one has $\lambda
_{\beta }\sim 1/L$. For small values of $L$ and for the nonzero modes, $\lambda
_{\beta }$ is large. The contribution of these modes is exponentially
suppressed and the main contribution comes form the zero mode. In this case,
from (\ref{P1}) we recover the Casimir force for two parallel plates in $%
(D_{1}+1)$-dimensional Minkowskian spacetime, namely%
\begin{equation}
V_{\Sigma }P\approx -\frac{2(4\pi )^{-D_{1}/2}}{\Gamma (D_{1}/2)}%
\int_{m}^{\infty }dx\,\frac{x^{2}(x^{2}-m^{2})\,^{D_{1}/2-1}}{\frac{(\beta
_{1}x-1)(\beta _{2}x-1)}{(\beta _{1}x+1)(\beta _{2}x+1)}e^{2ax}-1}.
\label{ForceD1}
\end{equation}%
For the case of a degenerated zero eigenstate the corresponding degeneracy
factor must be included on the right-hand side. In some models of
compactification the zero mode is absent (for example, in models with
twisted boundary conditions along the compactified dimensions, see below).
In such cases, for small values of $L$ the main contribution to the Casimir
force comes from the lowest mode $\lambda _{\beta }=\lambda _{0}$ and, to
leading order, one gets%
\begin{equation}
V_{\Sigma }P\approx -\frac{m_{0}^{D_{1}+1}}{(4\pi )^{D_{1}/2}}\frac{(\beta
_{1}m_{0}x+1)(\beta _{2}m_{0}x+1)}{(\beta _{1}m_{0}x-1)(\beta _{2}m_{0}x-1)}%
\frac{e^{-2am_{0}}}{(am_{0})^{D_{1}/2}},  \label{Psmall}
\end{equation}%
where $m_{0}=\sqrt{\lambda _{0}^{2}+m^{2}}$. Hence, here the Casimir
forces are exponentially suppressed for small sizes of the internal space.

For small values of the inter-plate distance, $a/|\beta _{j}|\ll 1$, the main
contribution into the integral in Eq.~(\ref{P1}) comes from larger values of $x$
and, to leading order, one has%
\begin{equation}
P\approx -\frac{2(4\pi )^{-D_{1}/2}}{V_{\Sigma }\Gamma (D_{1}/2)}\sum_{\beta
}\int_{m_{\beta }}^{\infty }dx\,x^{2}\frac{(x^{2}-m_{\beta
}^{2})\,^{D_{1}/2-1}}{e^{2ax}-1},  \label{ForceSm}
\end{equation}%
except for the case of Dirichlet boundary conditions on one plate and
non-Dirichlet boundary conditions on the other.
We see that in this limit (\ref{ForceSm}) the Casimir
force is attractive. However, in the case of Dirichlet boundary condition on one
plate and non-Dirichlet boundary condition on the other the Casimir force
is repulsive at small distances, what is indeed a remarkable result.

\section{Particular cases}

\label{sec:example}

As a simple example of a particular application of the general
results obtained above, we will first
consider the special case where $\Sigma =S^{1}$, with the size of the internal
space being $2\pi L$. For the compact dimension we assume a
general periodicity condition of the form%
\begin{equation}
\psi _{\beta }(X+2\pi L)=e^{2\pi i\alpha }\psi _{\beta }(X),  \label{PerCond}
\end{equation}%
with constant $\alpha $, $0\leqslant \alpha \leqslant 1$. The specific cases $%
\alpha =0$ and $\alpha =1/2$ correspond to untwisted and to twisted fields, respectively. The
corresponding part of the scalar eigenfunctions is%
\begin{equation}
\psi _{\beta }(X)=\frac{e^{iKX}}{\sqrt{2\pi L}},\;K=(\beta +\alpha
)/L,\;\beta =0,\pm 1,\pm 2,\ldots .  \label{psibetS1}
\end{equation}%
The formulas for the topology $R^{(D-1,1)}\times S^{1}$ are obtained
from the results given in the previous sections, by taking%
\begin{equation}
\sum_{\beta }=\sum_{\beta =-\infty }^{+\infty },\;\lambda _{\beta }=\frac{%
|\beta +\alpha |}{L},\;m_{\beta }=\sqrt{(\beta +\alpha )^{2}/L^{2}+m^{2}}%
,\;D_{1}=D-1.  \label{Changes}
\end{equation}%
In particular, for the Casimir force one has%
\begin{equation}
P=-\frac{(4\pi )^{-(D-1)/2}}{\pi \Gamma ((D-1)/2)L}\sum_{\beta =-\infty
}^{+\infty }\int_{m_{\beta }}^{\infty }dx\,\frac{x^{2}(x^{2}-m_{\beta
}^{2})\,^{(D-3)/2}}{\frac{(\beta _{1}x-1)(\beta _{2}x-1)}{(\beta
_{1}x+1)(\beta _{2}x+1)}e^{2ax}-1}.  \label{PS1}
\end{equation}
For large values of $L$ the main contribution to the series in (\ref%
{PS1}) comes from large values of $\beta $ and one can replace the summation
over $\beta $ by an integration. After some transformations, to leading
order we find the result%
\begin{equation}
P\approx -\frac{2(4\pi )^{-D/2}}{\Gamma (D/2)}\int_{0}^{\infty }dx\frac{%
x^{2}(x^{2}-m^{2})^{D/2-1}}{\frac{(\beta _{1}x-1)(\beta _{2}x-1)}{(\beta
_{1}x+1)(\beta _{2}x+1)}e^{2ax}-1},  \label{PLlarge}
\end{equation}%
which is, in fact, the Casimir force for two parallel plates in $(D+1)$-dimensional
Minkowskian spacetime.

\begin{figure}[tbph]
\begin{center}
\epsfig{figure=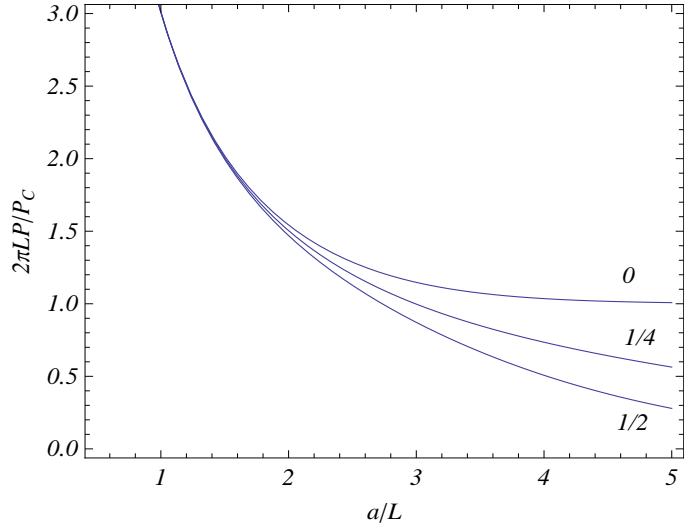,width=9.cm,height=7.cm}
\end{center}
\caption{Ratio of the Casimir force for two parallel plates in the
spacetime with topology $R^{(3,1)}\times S^{1}$ to the standard Casimir
force in $R^{(3,1)}$, for a massless Dirichlet scalar, as a function of $a/L$.
The values on each of the curves correspond to those of the parameter $%
\protect\alpha $.}
\label{fig2}
\end{figure}

As we already noted before, depending on the values of the
coefficients $\beta _{j}$ and of the distance between the plates, the Casimir
force (\ref{PS1}) can be either attractive or repulsive. In Fig.~\ref{fig2},
and corresponding to the model with $D=4$ and for a massless scalar field with Dirichlet
boundary conditions, we have plotted the ratio $2\pi LP/P_{C}$ as a function
of $a/L$, where $P_{C}=-\pi ^{2}/(480a^{4})$ is the standard Casimir force.
The values on each of the curves correspond to those of the parameter $%
\alpha $. As we have explained before, for $\alpha \neq 0$ the zero mode is
absent and for large values of $a/L$ the Casimir force is exponentially
suppressed.

\begin{figure}[tbph]
\begin{center}
\epsfig{figure=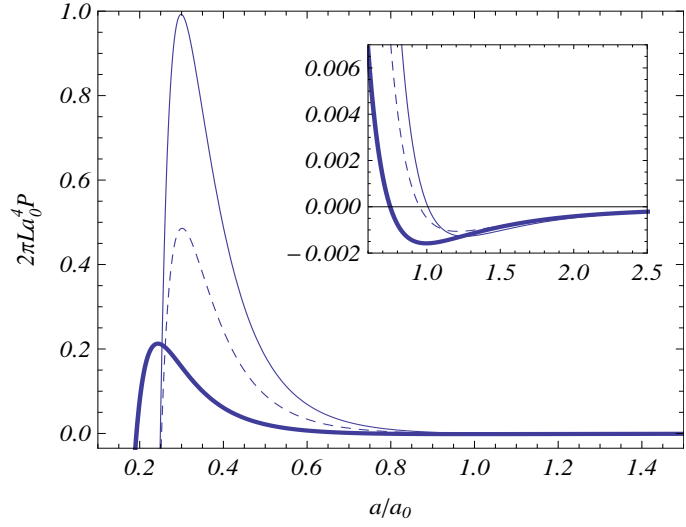,width=9.cm,height=7.cm}
\end{center}
\caption{Casimir force for two parallel plates in the
spacetime with topology $R^{(3,1)}\times S^{1}$, for an untwisted
massless scalar field with
Robin coefficients $\protect\beta _{1}/a_{0}=-0.1$, $\protect\beta %
_{2}/a_{0}=-0.5$, as a function of the distance between the plates.
The full (dashed) curve corresponds to a size of the internal space with $L/a_{0}=1$ ($%
L/a_{0}=0.5$). The thick curve corresponds to the Casimir force
for two parallel plates in Minkowski spacetime $R^{(3,1)}$
with the same Robin coefficients.} \label{fig3}
\end{figure}

In Fig.~\ref{fig3} the Casimir force is plotted for the topology $%
R^{(3,1)}\times S^{1}$ and an untwisted ($\alpha =0$) massless
scalar field with Robin coefficients $\beta _{1}/a_{0}=-0.1$, $\beta
_{2}/a_{0}=-0.5$ ($a_{0}$ a fixed length scale) as a function of $%
a/a_{0}$ for $L/a_{0}=1$ (full curve) and $L/a_{0}=0.5$ (dashed
curve). The thick curve corresponds to the Casimir force for two
parallel plates in Minkowski spacetime $R^{(3,1)}$ with the
same Robin coefficients. As is seen, the corresponding Casimir
forces are attractive for small and large distances between the
plates while they are repulsive for intermediate distances. There
are two equilibrium points corresponding to the zeros of the
function $P$. The leftmost point is unstable whereas the rightmost one is
stable. Hence, in this case the Casimir force {\it stabilizes} the
distance between the plates. This feature can be used in
braneworld models for the stabilization of the radion field. We see
 from Fig.~\ref{fig3} that the height of the barrier between
the stable and the unstable equilibrium points is increased by the
presence of the internal space. As a consequence, an
enhancement of the repulsive Casimir effect, coming from the extra dimension, occurs.

\begin{figure}[tbph]
\begin{center}
\epsfig{figure=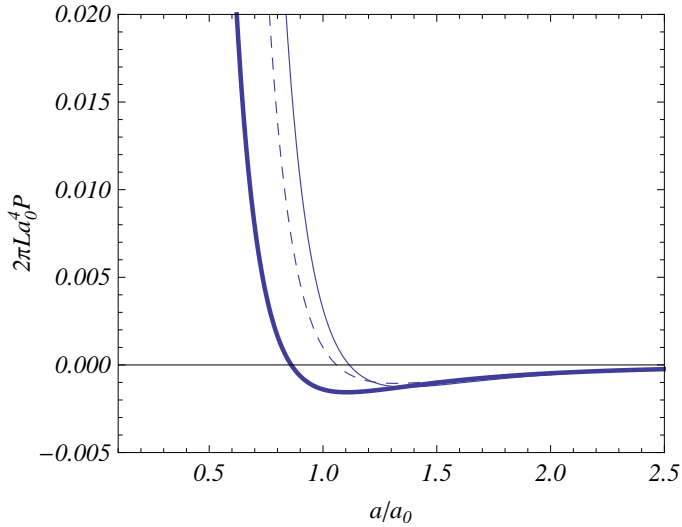,width=9.cm,height=7.cm}
\end{center}
\caption{Same as in Fig.~\ref{fig3} for a scalar field with
Robin coefficients $\protect\beta _{1}=0$, $\protect\beta %
_{2}/a_{0}=-0.5$.} \label{fig4}
\end{figure}

As already explained before, in the case of Dirichlet boundary
conditions on one plate and non-Dirichlet boundary conditions on the
other, the Casimir force is repulsive at small separations. In
Fig.~\ref{fig4} we illustrate this feature for the topology
$R^{(3,1)}\times S^{1}$ in the case of an untwisted massless
scalar field with $\beta _{1}=0$, $\beta
_{2}/a_{0}=-0.5$. As in Fig.~\ref{fig3}, the full (dashed) curve stands for  $%
L/a_{0}=1$ ($L/a_{0}=0.5$) and the thick curve corresponds to the
Casimir force for two parallel plates in  Minkowski spacetime
$R^{(3,1)}$ with the same Robin coefficients.

For the topology under consideration, the Casimir energy for the
bulk without boundaries,%
\begin{equation*}
E_{R^{(D-1,1)}\times S^{1}}=\frac{1}{2}\int \frac{d\mathbf{k}_{D-1}}{(2\pi
)^{D-1}}\sum_{\beta =-\infty }^{+\infty }\sqrt{k_{D-1}^{2}+(\beta +\alpha
)^{2}/L^{2}+m^{2}}
\end{equation*}%
can be further simplified through the Abel-Plana
summation formula, in the form \cite{SahRev07,Beze08}%
\begin{equation}
\sum_{\beta =-\infty }^{+\infty }f(|\beta +\alpha |)=2\int_{0}^{\infty
}dx\,f(x)+i\int_{0}^{\infty }dx\,\sum_{\lambda =\pm 1}\frac{f(ix)-f(-ix)}{%
e^{2\pi (x+i\lambda \alpha )}-1},  \label{sumform1}
\end{equation}%
what leads to the result%
\begin{eqnarray}
E_{R^{(D-1,1)}\times S^{1}} &=&\frac{2\pi L}{2}\int \frac{d^{D}k}{(2\pi )^{D}%
}\sqrt{k_{D}^{2}+m^{2}}  \notag \\
&&-\frac{2(Lm)^{(D+1)/2}}{(2\pi )^{D}L^{D}}\sum_{n=1}^{\infty }\frac{\cos
(2\pi n\alpha )}{n^{(D+1)/2}}K_{(D+1)/2}(2\pi nLm).  \label{EL2}
\end{eqnarray}%
The second term on the right-hand side of this expression is finite and
introduction of a cutoff function is necessary for the first term only.
Note that the latter is the vacuum energy density for the spatial topology $%
R^{D}$ and, hence, the second term on the right-hand side of Eq.~(\ref{EL2})
is the contribution to the vacuum energy induced by the compactness of the $x^{D}$
dimension. In particular, the topological part of the vacuum energy is
always negative (positive) for untwisted (twisted) scalars.

In a similar way, we can also extract the topological contributions in the single
plate terms of the vacuum energy. After applying the summation
formula (\ref{sumform1}) and after integration, these contributions yield%
\begin{eqnarray}
E_{j} &=&2\pi LE_{j}^{(M)}+\frac{(Lm)^{D/2}}{2(2\pi L)^{D-1}}%
\sum_{n=1}^{\infty }\frac{\cos (2\pi n\alpha )}{n^{D/2}}K_{D/2}(2\pi nLm)
\notag \\
&&+\frac{(4\pi )^{-(D-1)/2}L\beta _{j}}{\Gamma ((D+1)/2)}\sum_{\lambda =\pm
1}\int_{m}^{\infty }dy\frac{(y^{2}-m^{2})^{D/2}}{e^{2\pi Ly+2\pi i\lambda
\alpha }-1}\int_{0}^{1}dx\frac{(1-x^{2})^{(D-1)/2}}{1+\beta
_{j}^{2}(y^{2}-m^{2})x^{2}},  \label{EjS1}
\end{eqnarray}%
where
\begin{equation}
E_{j}^{(M)}=-\frac{1}{8}\int \frac{d\mathbf{k}_{D-1}}{(2\pi )^{D-1}}\sqrt{%
k_{D-1}^{2}+m^{2}}-\frac{\beta _{j}}{2\pi }\int \frac{d\mathbf{k}_{D-1}}{%
(2\pi )^{D-1}}\int_{0}^{\infty }dx\frac{\sqrt{k_{D-1}^{2}+x^{2}+m^{2}}}{%
1+\beta _{j}^{2}x^{2}}  \label{EjM}
\end{equation}%
is the vacuum energy (per unit volume along the coordinates $x^{1},\ldots
,x^{D-1}$) for a single plate in Minkowski spacetime with trivial topology $%
R^{(D,1)}$. Hence, the last two terms on the right-hand side of (\ref{EjS1})
are the terms in the vacuum energy corresponding to a single plate and due to the compactness
of the dimension $x^{D}$. These terms are finite and renormalization is
needed for the Minowskian part $E_{j}^{(M)}$ only. Note that for Dirichlet
and Neumann boundary conditions the last term on the right of Eq.~(\ref{EjS1}) vanishes.

The case of a $D_{2}$-dimensional torus as internal space, $\Sigma
=(S^{1})^{D_{2}}$, can be considered in a similar way. For a scalar field
with the periodicity condition $\psi _{\beta }(X^{l}+2\pi L_{l})=e^{2\pi
i\alpha _{l}}\psi _{\beta }(X^{l})$ along the coordinate $X^{l}$, $%
0\leqslant X^{l}\leqslant 2\pi L_{l}$, the formulas for the Casimir energy
and force are obtained from the general expressions in Sects.~\ref%
{sec:CasEn} and \ref{sec:CasForce} with the substitutions%
\begin{equation}
\sum_{\beta }=\sum_{j_{1}=-\infty }^{+\infty }\cdots \sum_{j_{D_{2}}=-\infty
}^{+\infty },\;\;m_{\beta }^{2}=\sum_{l=1}^{D_{2}}(j_{l}+\alpha
_{l})^{2}/L_{l}^{2}+m^{2}.  \label{torus}
\end{equation}

Concerning the issues of embedding the model in string theory and
of the discussions of the holographic principle there, the case of the
internal space $\Sigma =S^{D_{2}}$ is of very special interest. The corresponding
eigenfunctions $\psi _{\beta }(X)$ are expressed in terms of spherical
harmonics of degree $l$, $l=0,1,2,\ldots $. For the internal space with
radius $L$ the expressions for the Casimir energy and Casimir force are
quite easily obtained from the general formulas given in the above
sections, just by replacing%
\begin{eqnarray}
\sum_{\beta } &\rightarrow &\sum_{l=0}^{\infty }(2l+D_{2}-1)\frac{\Gamma
(l+D_{2}-1)}{l!\Gamma (D_{2})},  \notag \\
\lambda _{\beta } &\rightarrow &\frac{1}{L}\sqrt{l(l+D_{2}-1)+\zeta
D_{2}(D_{2}-1)}.  \label{SD2}
\end{eqnarray}%
Here the factor under the summation sign is the degeneracy of the angular
mode with a given $l$.

\section{Generalized piston geometry}

\label{sec:Piston}

In a way very much similar to the procedure described in the
preceding sections, we are able to treat the more general case
when a part of the dimensions $x^{1},\ldots ,x^{D_{1}-1}$ are
still constrained by boundary conditions. This corresponds to
considering a generalized piston geometry, a quite fashionable
situation nowadays, in particular, in the quest for negative
Casimir forces (for the investigation of the Casimir effect in a
piston geometry see \cite{Cava04} and references therein). Here we
have obtained those in the configurations above, but would like to
see now the differences introduced in our results by the
consideration of piston geometries.

We will denote by $d_{1}$ the number of unconstrained
dimensions (coordinates $x^{1},\ldots ,x^{d_{1}}$) and by $\gamma _{i}^{2}$,
with a collective index $i$, the eigenvalues of the Laplacian along the
constrained directions:%
\begin{equation}
\Delta _{D_{1}-1-d_{1}}\varphi _{\alpha }(x^{M})=-\gamma _{i}^{2}\varphi
_{\alpha }(x^{M}).  \label{LapConstr}
\end{equation}%
The eigenfrequencies in the region between the plates are here given by
\begin{equation}
\omega (z_{n})=\sqrt{k_{d_{1}}^{2}+z_{n}^{2}/a^{2}+\gamma _{i}^{2}+m_{\beta
}^{2}},  \label{ompiston}
\end{equation}%
with the vacuum energy being%
\begin{equation}
E_{[a_{1},a_{2}]}=\frac{1}{2}\int \frac{d\mathbf{k}_{d_{1}}}{(2\pi )^{d_{1}}}%
\sum_{i,\beta }\sum_{n=1}^{\infty }\sqrt{k_{d_{1}}^{2}+z_{n}^{2}/a^{2}+%
\gamma _{i}^{2}+m_{\beta }^{2}}.  \label{Epist}
\end{equation}%
After applying the summation formula (\ref{sumfor}) to the sum over $n$,
we can write the energy in the decomposed form%
\begin{equation}
E_{[a_{1},a_{2}]}=\frac{a}{2}\int \frac{d\mathbf{k}_{d_{1}+1}}{(2\pi
)^{d_{1}+1}}\sum_{i,\beta }\sqrt{k_{d_{1}+1}^{2}+\gamma _{i}^{2}+m_{\beta
}^{2}}+\sum_{j=1,2}E_{j}+\Delta E_{[a_{1},a_{2}]}.  \label{EdecPist}
\end{equation}%
Here the expressions for the terms $E_{j}$ and $\Delta E_{[a_{1},a_{2}]}$
are obtained from Eqs.~(\ref{Ej}) and (\ref{DelE0}) by the replacements%
\begin{eqnarray}
&&\int \frac{d\mathbf{k}}{(2\pi )^{D_{1}-1}}\rightarrow \int \frac{d\mathbf{k%
}_{d_{1}}}{(2\pi )^{d_{1}}},\;k^{2}\rightarrow k_{d_{1}}^{2},
\label{Replace} \\
&&\sum_{\beta }\rightarrow \sum_{i,\beta },\;D_{1}\rightarrow
d_{1}+1,\;m_{\beta }^{2}\rightarrow \gamma _{i}^{2}+m_{\beta }^{2}.
\label{Replace2}
\end{eqnarray}%
These formulas are further simplified after integrating over the angular
part of the vector $\mathbf{k}_{d_{1}}$. \ The corresponding expressions are
obtained from the results of Sects.~\ref{sec:CasEn} and \ref{sec:CasForce},
with the replacements (\ref{Replace2}). In particular, for the interaction
part of the Casimir energy (per unit volume along the direction $%
x^{1},\ldots ,x^{d_{1}}$) one has%
\begin{eqnarray}
\Delta E_{[a_{1},a_{2}]} &=&\frac{(4\pi )^{-(d_{1}+1)/2}}{\Gamma
((d_{1}+1)/2)}\sum_{i,\beta }\int_{\sqrt{\gamma _{i}^{2}+m_{\beta }^{2}}%
}^{\infty }dx\,x(x^{2}-\gamma _{i}^{2}-m_{\beta }^{2})\,^{(d_{1}-1)/2}
\notag \\
&&\times \ln \left[ 1-\frac{(\beta _{1}x+1)(\beta _{2}x+1)}{(\beta
_{1}x-1)(\beta _{2}x-1)}e^{-2ax}\right] \,.  \label{DelEPist}
\end{eqnarray}%
The expression for the Casimir pressure takes the form
\begin{equation}
P(a,\beta _{1},\beta _{2})=-\frac{2(4\pi )^{-(d_{1}+1)/2}}{V_{\mathrm{cs}%
}V_{\Sigma }\Gamma ((d_{1}+1)/2)}\sum_{i,\beta }\int_{\sqrt{\gamma
_{i}^{2}+m_{\beta }^{2}}}^{\infty }dx\,\frac{x^{2}(x^{2}-\gamma
_{i}^{2}-m_{\beta }^{2})\,^{(d_{1}-1)/2}}{\frac{(\beta _{1}x-1)(\beta
_{2}x-1)}{(\beta _{1}x+1)(\beta _{2}x+1)}e^{2ax}-1},  \label{PCasPist}
\end{equation}%
where $V_{\mathrm{cs}}$ is the volume of the piston cross section along
the coordinates $x^{d_{1}+1},\ldots ,x^{D_{1}-1}$. In particular, for Dirichlet
and Neumann boundary conditions on the plates, we find%
\begin{equation}
P^{\mathrm{(J,J)}}(a)=\frac{2a^{-d_{1}-2}}{(8\pi )^{d_{1}/2+1}V_{\mathrm{cs}%
}V_{\Sigma }}\sum_{i,\beta }\sum_{n=1}^{\infty }\frac{1}{n^{d_{1}+2}}\left[
f_{d_{1}/2+1}(z)-f_{d_{1}/2+2}(z)\right] _{z=2na\sqrt{\gamma
_{i}^{2}+m_{\beta }^{2}}},  \label{PJJpist}
\end{equation}%
$\mathrm{J=D,N}$, the function $f_{\nu }(z)$ being defined in Eq.~(\ref{fnu}).
The corresponding force remains attractive independently of the form of
the cross section. In the case of Dirichlet boundary condition on one plate
and Neumann boundary condition on the other the expression for the Casimir
force is obtained from (\ref{PJJpist}) after introducing an additional factor $%
(-1)^{n+1}$, and the resulting force is always repulsive. In the special case
$d_{1}=0$ and for a massless scalar field, Eq.~(\ref{PJJpist}) reduces to
the formula for the Casimir force in Ref.~\cite{Full09b}, as it should.

On the base of Eqs.~(\ref{DelEPist}) and (\ref{PCasPist}) we can analyze the
geometry of a generalized piston with two chambers, assuming that the plates
are located at $x^{D_{1}}=a_{0},a_{1},a_{2}$, with the Robin coefficient $%
\beta _{0}$ for the left plate (for the Casimir effect in a piston geometry
see, for example, Refs.~\cite{Cava04}). For an arbitrary cross section, the
effective pressure on the plate at $x^{D_{1}}=a_{1}$ is given by
\begin{equation}
P_{a_{1}}=P(a,\beta _{1},\beta _{2})-P(b,\beta _{0},\beta _{1}),
\label{Pa1pist}
\end{equation}%
with $b=a_{1}-a_{0}$. For $P_{a_{1}}<0$ ($P_{a_{1}}>0$) the resulting force
on the plate is directed towards the right (left) plate. For Dirichlet and
Neumann boundary conditions the Casimir stress given by Eq.~(\ref{PJJpist})
is a monotonic function of the plate separation and, hence, in the piston
geometry with two chambers the resulting force (\ref{Pa1pist}) is directed
toward the closer plate.

In the special case of the geometry of a piston with circular
cross section in the plane $(x^{D_{1}-2},x^{D_{1}-1})$, the
corresponding part in the eigenfunctions has the form
$J_{|q|}(\eta r)e^{iq\phi }$, $q=0,\pm 1,\pm 2$, where $r$ and
$\phi $ are polar coordinates on this plane and $J_{q}(z)$ is the
Bessel function. The eigenvalues for the quantum number $\eta $
are quantized by the boundary conditions on the cylindrical
surface $r=r_{0}$, with $r_{0}$ being the piston radius. For
example, in the case of a Dirichlet boundary condition one has
$\eta =j_{|q|,p}/r_{0}$, where $j_{\nu ,p}$ is the $p$-th positive
zero of the function $J_{\nu }(z)$. The corresponding formulas for
the Casimir energy and forces are obtained from the general
results (\ref{DelEPist}) and (\ref{PCas}) with the substitutions%
\begin{equation}
d_{1}=D_{1}-3,\;\sum_{i}=\sum_{q=-\infty }^{+\infty }\sum_{p=1}^{\infty
},\;\gamma _{i}=j_{|q|,p}/r_{0}.  \label{Replace3}
\end{equation}%
Other types of boundary conditions on the cylindrical boundary
can be considered in a similar way. Moreover, the generalization of our procedure
to more than three plates is also easy to carry out.

\section{Conclusion}

\label{sec:Conc}

We have investigated in this paper the influence of extra
dimensions on the Casimir energy and on the Casimir force for a
massive scalar field with an arbitrary curvature coupling
parameter, in the usual geometry of two parallel plates. We have
assumed that on the plates the field obeys Robin boundary
conditions with, in general, different coefficients for the two
different plates. The corresponding eigenfrequencies are expressed
in terms of solutions of a transcendental equation (\ref{eigeq}),
thus they are known implicitly only. By applying the summation
formula (\ref{sumfor}) to the corresponding series in the mode-sum
for the vacuum energy in the region between the plates, we have
explicitly extracted, in a cut-off independent way, the
boundary-free (topological) part and the contributions induced by
the single plates (when the other plate is absent). The remaining
interaction part is finite for all nonzero inter-plate distances
and is cut-off independent. The surface divergences in the Casimir
energy are contained in the single plate components only. But the
latter do not depend on the location of the plate and do not
contribute to the Casimir force. For an arbitrary internal space,
the interaction part of the Casimir energy is given by
Eq.~(\ref{DelE}). In the special cases of Dirichlet and Neumann
boundary conditions on both plates this formula leads to the
result (\ref{DELEDN1}) and the corresponding energy is always
negative. For Dirichlet boundary conditions on one plate and
Neumann boundary conditions on the other, the interaction
component of the vacuum energy is given by Eq.~(\ref{DelED+N}),
and it is positive for all values of the interplate distance. In
the case of a conformally coupled massless field on the background
of a spacetime conformally related to the one
described by the line element (\ref{metric}) with the conformal factor $%
\Omega ^{2}(x^{D_{1}})$, the interaction part of the Casimir
energy is given by Eq.~(\ref{DelE}), with the coefficients $\beta
_{j}$ being related to the specific coefficients of the Robin
boundary conditions (\ref{BCbrane}) and to the conformal factor by
Eqs.~(\ref{relbeta}). In the Randall-Sundrum two brane model with
a compact internal space, the corresponding Robin coefficients are
given by Eq.~(\ref{betajRS}) and the corresponding vacuum energy
can have a minimum, corresponding to the stable equilibrium point.
This feature is useful in braneworld models for the stabilization
of the radion field.

The interaction forces between the plates for the most general
case of internal space have been considered in
Sect.~\ref{sec:CasForce}. In order to obtain the resulting force,
the contributions from both sides of the plates must be taken into
account. Then, the forces coming from the topological parts of the
vacuum energy cancel out and only the interaction terms contribute
to the Casimir force. In order to show this important fact
explicitly, we have considered a piston-like geometry, by
introducing a third plate. At the end of the calculation this
plate is sent to infinity. The resulting Casimir force is given by
Eq.~(\ref{PCas}). With independence of the geometry of the
internal space, the force is attractive for Dirichlet or Neumann
boundary conditions on both plates (formula (\ref{FDN})) and it is
repulsive for Dirichlet boundary conditions on one plate and
Neumann boundary conditions on the other. In both cases the force
is a monotonic function of the distance. For general Robin
boundary conditions the Casimir force can be either attractive
(corresponding to negative values of $P$) or repulsive (positive
values of $P$), depending on the particular Robin coefficients and
on the distance between the plates. For small values of the size
of the internal space and in models where the zero modes along the
internal space are present, the main contribution to the Casimir
force comes from the zero modes and the contributions of the
nonzero modes are exponentially suppressed. In this limit, to
leading order we recover the standard result for the Casimir force
between two plates in $(D_{1}+1)$-dimensional Minkowski spacetime.
When the zero mode is absent (for example, in the case of twisted
boundary conditions along the compactified dimensions), the
Casimir forces are exponentially suppressed in the limit of small
size of the internal space. For small values of the inter-plate
distance the Casimir forces are attractive, independently of the
values of the Robin coefficients, except for the case of Dirichlet
boundary conditions on one plate and non-Dirichlet boundary
conditions on the other. In this latter case, the Casimir force is
repulsive at small distances. It is interesting to remark that
this property could be used in the proposal of a Casimir
experiment with the purpose to carry out an explicit detailed
observation of `large' extra dimensions as allowed by some models
of particle physics.

As an illustration of the general results, in
Sect.~\ref{sec:example} we have considered a special model for the
internal space $\Sigma =S^{1}$, with the periodicity condition
(\ref{PerCond}) along the compactified dimension. For the specific
values $\alpha =0,1/2$ this condition corresponds to untwisted and
twisted scalar fields, respectively. In Fig.~\ref{fig2}, for
Dirichlet boundary conditions we depicted the dependence of the
Casimir force on
the distance between the plates for different values of the parameter $%
\alpha $. In Fig.~\ref{fig3} a plot of the Casimir force in the
case of Robin boundary conditions on both plates as a function of
the inter-plate distance has been provided. In the example
considered, the Casimir force is attractive both for large and for
small distances, while it is repulsive at intermediate distances.
The Casimir force vanishes at two values of the inter-plate
distance, which correspond to equilibrium points. The leftmost
point is unstable and the rightmost one is locally stable. As
shown in the plot, the stability of the rightmost equilibrium
point is enhanced by the presence of the internal space. Formulas
for the Casimir energy and force for the more general internal
spaces $(S^{1})^{D_{2}}$ ($D_{2}$-torus) and $S^{D_{2}}$ have been
obtained from the general results of Sects.~\ref{sec:CasEn} and~\ref%
{sec:CasForce}.

In the last section \ref{sec:Piston}, we have extended the results
from the previous ones to the case of a piston geometry with
finite cross section of arbitrary form along some subset of the
dimensions. The corresponding expressions for the Casimir energy
and Casimir force between the plates are given by
Eqs.~(\ref{DelEPist}) and (\ref{PCasPist}). We have checked that
the qualitative features described above remain basically
unaltered. In particular, the possibility of a repulsive Casimir
effect is again observed. In the special case of a piston geometry
of circular cross section on the plane
$(x^{D_{1}-2},x^{D_{1}-1})$, the corresponding formulas have been
specified in (\ref{Replace3}).

The search for specific applications of this study to practical
situations in braneworld models, nano-physics and particle physics
will keep us busy for some time.

\section{Acknowledgments}

E.E. has been supported by MICIIN (Spain), project FIS2006-02842,
and by AGAUR (Generalitat de  Ca\-ta\-lu\-nya), contract
2005SGR-00790 and grant DGR2008BE1-00180. S.D.O. was supported in
part by projects FIS2006-02842 and PIE2007-50I023 (MICIIN, Spain).
A.A.S. was supported by the ESF Programme ``New Trends and
Applications of the Casimir Effect" and in part by the Armenian
Ministry of Education and Science Grant No. 119. A.A.S. gratefully
acknowledges the hospitality of the Abdus Salam International
Centre for Theoretical Physics (Trieste, Italy) where part of this
work was done.

\end{document}